 \documentclass[12pt,preprint]{aastex}

\shorttitle{Kappa Distribution For Solar Flare HXR Source}
\shortauthors{Oka et al.}

\begin{document}

\title{
Kappa Distribution Model for Hard X-Ray Coronal Sources of Solar Flares
}


\author{M. Oka\altaffilmark{1}, S. Ishikawa\altaffilmark{1,2},  P. Saint-Hilaire\altaffilmark{1}, S. Krucker\altaffilmark{1,3}, R. P. Lin\altaffilmark{1,4}}

\altaffiltext{1}{Space Sciences Laboratory, University of California Berkeley, USA}
\altaffiltext{2}{National Astronomical Observatory of Japan, Japan}
\altaffiltext{3}{i4Ds, University of Applied Sciences and Arts Northwestern Switzerland, Switzerland}
\altaffiltext{4}{School of Space Research, Kyung Hee University, Republic of Korea}

%
%
%

%


\begin{abstract}
Solar flares produce hard X-ray emission of which the photon spectrum is often represented by a combination of thermal and power-law distributions. However, the estimates of the number and total energy of non-thermal electrons are sensitive to the determination of the power-law cutoff energy. Here we revisit an `above-the-loop' coronal source observed by RHESSI on 2007 December 31 and show that a kappa distribution model can also be used to fit its spectrum. Because the kappa distribution has a Maxwellian-like core in addition to the high-energy power-law tail, the emission measure and temperature of the instantaneous electrons can be derived without assuming the cutoff energy. Moreover, the non-thermal fractions of electron number/energy densities can be uniquely estimated because they are functions of the power-law index only. With the kappa distribution model, we estimated that the total electron density of the coronal source region was $\sim$ 2.4 $\times$ 10$^{10}$ cm$^{-3}$. We also estimated without assuming the source volume that a moderate fraction ($\sim$ 20\%) of electrons in the source region was non-thermal and carried $\sim$ 52\% of the total electron energy. The temperature was 28 MK, and the power-law index $\delta$ of the electron density distribution was  -4.3. These results are compared to the conventional power-law models with and without a thermal core component.



\end{abstract}





\section{Introduction}

A solar flare is an explosive energy release phenomenon on the Sun and accelerates a large number of electrons up to tens of MeV \citep[e.g.][]{brown71, lin76, miller97, holman03L}.  To diagnose accelerated electrons, the hard X-ray (HXR) observations of electron bremsstrahlung emission have been used. 

In general, the spatially integrated HXR photon spectrum exhibits a relatively flat, non-thermal tail in addition to an intense and steep thermal component \citep{lin81}. 
Although a model with multiple temperatures can often fit the entire spectra \citep{emslie80}, the non-thermal tail can typically be described as a power-law or double power-law that connects with the thermal component typically in the 15-30 keV range. 

When viewed as an image, the intense thermal emission is dominated by an arcade loop structure whereas the less intense but the high energy tail of the HXR emission is usually detected at chromospheric footpoint of the loop \citep[e.g.][]{hoyng81, brown83}. The non-thermal tail can also originate from the corona \citep[e.g.][]{frost71, palmer72}, and the source is sometimes located  `above-the-loop'  \citep[e.g.][]{masuda94, krucker08, ishikawa11}.

A caveat of studying the non-thermal HXR emission is that, the thermal emission from a loop is so bright that it masks spectral features of the non-thermal sources (in either footpoint or `above-the-loop') especially in the lower energy range. Therefore, it is difficult to clarify how far the power-law spectrum extends in the lower energy direction. As such, a low energy cutoff $E_{\rm c}$ of the power-law has been considered, typically in the 15-30 keV range, to estimate the number and total energy of non-thermal electrons in the source, although the estimates can be sensitive to the choice of $E_{\rm c}$.

Thus, efforts have been made to understand properties of HXR emission around $E_{\rm c}$ \citep[e.g.][]{holman92, sui05}. In particular, since the launch of RHESSI, it has been successfully shown that a range of values for $E_{\rm c}$ fit the data equally well and that the highest value of $E_{\rm c}$ that still fits the data can be used to derive the lower-limit for the non-thermal number and energy densities \citep{holman03L, emslie04, sainthilaire05, kontar08}. As for the physical meaning of the low-energy cutoff, it was argued that the cutoff represents the critical velocity above which electrons run away and are freely accelerated by the reconnection electric field \citep{holman92}. If a sharp cutoff existed, however, plasma instabilities would lead to flattening of the distribution around and below the cutoff energy \citep[as reviewed by][]{holman11}.

In fact, a theoretical study pointed out that the non-thermal electron distribution could seamlessly merge into a thermal distribution, removing the need for a low-energy cutoff \citep{emslie03}. Moreover, in-situ observations of electrons in the planetary and interplanetary space often show that the higher energy tail of a thermal core component smoothly extends into a power-law distribution. Examples can be found at the Earth's bow shock \citep[e.g.][]{gosling89} and the magnetotail reconnection \citep[e.g.][]{oieroset02}. 

In some cases of in-situ observations, the kappa distribution model \citep{vasyliunas68} has been used to represent the entire electron distribution because it is a composite of Maxwellian-like core and a power-law tail \citep[e.g.][]{christon88,christon89,christon91,onsager91,leubner04, imada11}.  While the kappa distribution was first introduced as an empirical model \citep{vasyliunas68}, recent theoretical and computational studies have suggested that self-consistent formation of electron kappa distribution is possible by the beam-plasma interactions which involve the Langmuir/ion-sound turbulence \citep{yoon06, rhee06, ryu07}. The origin of the kappa distribution has also been discussed in terms of Gibbsian theory \citep{treumann08} and Tsallis Statistical Mechanics \citep[][and references therein]{livadiotis11}.  From the solar physics point of view, \cite{kasparova09} has already suggested that the kappa distribution may also be useful for interpretations of solar HXR sources. They reported that a kappa distribution fits the spectrum of a coronal loop-top source but fits less well the spatially integrated spectrum (coronal and footpoint sources).

The purpose of this paper is to complement the work of \cite{kasparova09} by examining the kappa distribution model in  a recently reported RHESSI event of 2007 December 31 \citep{krucker10}. We studied this event because an unusually intense HXR emission was detected from `above-the-loop' coronal source.  We show that the spatially integrated HXR spectrum can be fitted by not only a combination of thermal and thin-target power-law distributions but also a combination of thermal and kappa distributions. The introduction of the core distribution in the non-thermal source via the kappa distribution enables us to estimate the number and energy densities without assuming the cutoff energy.

\section{Kappa Distribution}
\label{sec:kappa}


The isotropic, three-dimensional (3D) form of the kappa distribution function $f_\kappa(v)$ (s$^3$ cm$^{-6}$) is written as
\begin{equation}
f_\kappa(v) = \frac{N_\kappa}{(\pi\kappa\theta^2)^{3/2}} \frac{\Gamma(\kappa+1)}{\Gamma(\kappa-1/2)} 
\left(1+\frac{v^2}{\kappa \theta^2} \right)^{-(\kappa+1)}
\end{equation}
where $v$ is the particle speed, $\kappa$ is the power-law index, $\theta$ is the most probable particle speed, $\Gamma$ is the Gamma function and $N_\kappa$ is the number density. The coefficient is determined so that $\int f(v)d^3v$ = $N_\kappa$. If $\kappa$ is sufficiently large,
 the distribution approaches a single Maxwellian distribution. The most probable energy is $E_{\rm mp}$ = $(1/2)m\theta^2$ at which the differential flux (=$(v^2/m) f(v)$) becomes maximum. However, the temperature is defined as $k_BT_\kappa \equiv (1/2) m\theta^2 \left[ \kappa/(\kappa-3/2) \right]$ so that the average energy of particles can be expressed as $E_{\rm avg} = (3/2)k_BT_\kappa$. Note that $E_{\rm avg} = (3/2)k_BT_M$ for the isotropic 3D Maxwellian distribution $f_M(v)$ with $k_BT_M\equiv(1/2)mv_{th}^2$ where $v_{th}$ is the thermal speed.

By using the kappa temperature $k_BT_\kappa$ and introducing particle energy $E=(1/2)mv^2$, we can convert $f_\kappa(v)$ into the density distribution $F_\kappa(E)$ (cm$^{-3}$ keV$^{-1}$) as
\begin{eqnarray}
F_\kappa(E) &=& \frac{2N_\kappa\sqrt{E}}{\sqrt{\pi(k_BT_\kappa)^3}}\frac{\Gamma(\kappa+1)}
        {\left(\kappa-3/2 \right)^{3/2}\Gamma(\kappa-1/2)} \nonumber \\
     & & {\qquad} \times \left[1+\frac{E}{k_BT_\kappa(\kappa-3/2)} \right]^{-(\kappa+1)}
\end{eqnarray}
so that $\int F_\kappa(E)dE$ = $N_\kappa$. The thin-target formula of this expression has been incorporated into the Solar SoftWare (SSW) by \cite{kasparova09}  and can be used as OSPEX fitting function {\ttfamily f\_thin\_kappa.pro}.

\begin{figure}
\begin{center}
\includegraphics[width=90mm]{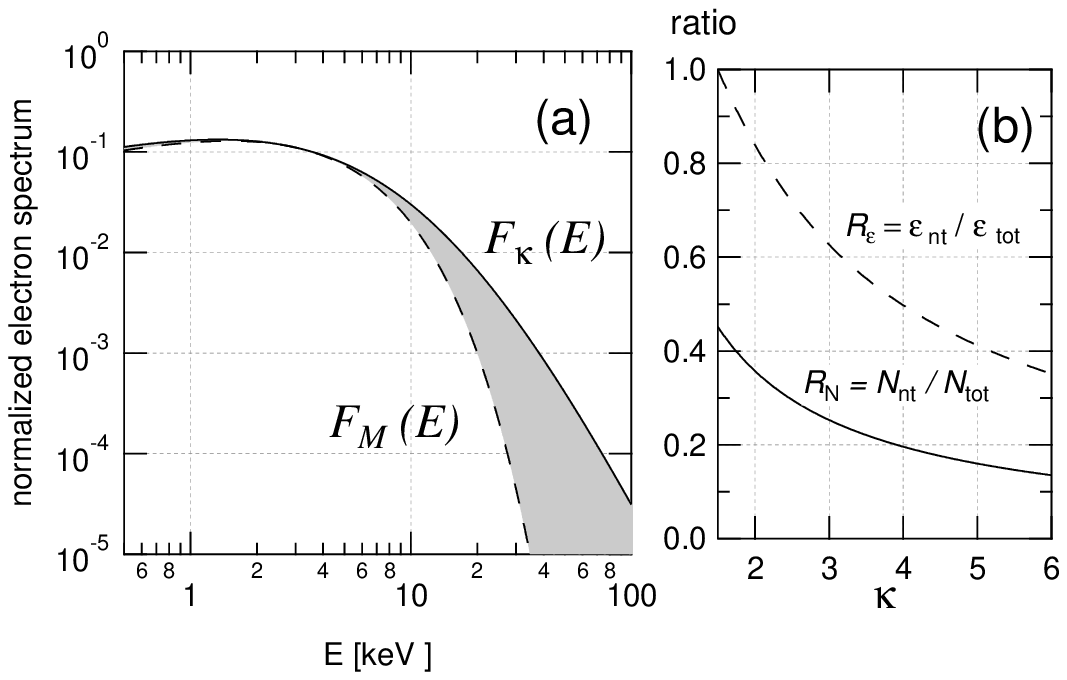}
\caption{
Properties of the kappa distribution $F_\kappa(E)$ compared with the adjusted Maxwellian distribution $F_M(E)$: (a) An example spectrum for $N_\kappa$ = 1.0, $\kappa$ = 4.0, $T_\kappa$=$E_{\rm mp}[\kappa/(\kappa-3/2)$], $N_M$ = 0.8 and $T_M$=$E_{\rm mp}$ where $E_{\rm mp}$ = 3.0 keV = 35 MK. (b) The density fraction $R_{\rm N}$ and the energy fraction $R_{\rm \varepsilon}$ of non-thermal electrons, plotted as functions of $\kappa$.
\label{fig:kappa}}
\end{center}
\end{figure}

An example of $F_\kappa(E)$ is plotted in Figure \ref{fig:kappa}(a). The Maxwellian distribution $F_M(E)$ with $kT_M = E_{\rm mp} = kT_\kappa[(\kappa-3/2)/\kappa]$ is also plotted for comparison. Here, the density $N_M$ of the Maxwellian distribution $F_M(E)$ has been adjusted so that $F_\kappa(E_{\rm mp}) = F_M(E_{\rm mp})$.  Such $N_M$ is derived as
\begin{equation}
\frac{N_M}{N_\kappa} = 2.718\frac{\Gamma(\kappa+1)}{\Gamma(\kappa-1/2)}\kappa^{-\frac{3}{2}}\left(1+\frac{1}{\kappa}\right)^{-(\kappa+1)}
\label{eq:ratio}
\end{equation}
Then, the difference between $F_\kappa(E)$ and the adjusted $F_M(E)$ represents the non-thermal particles, as indicated by the shaded region. A slight difference remains in the lower energy range ($E<E_{\rm mp}$) but the total difference in this range is negligible compared to the total difference in the higher energy range ($E>E_{\rm mp}$). Therefore, the ratio $R_N$ of the non-thermal electron density to the total electron density in a source can be approximated by $R_N \equiv 1 - N_M/N_k$. The ratio $R_\varepsilon$ of the non-thermal electron energy to the total electron energy can also be calculated using $E_{\rm avg}$. Both $R_N$ and $R_\varepsilon$ are functions of $\kappa$ only and are shown in Figure \ref{fig:kappa}(b). 

For the example case of $\kappa$ = 4 (i.e. Figure \ref{fig:kappa}(a)), the non-thermal electrons constitute $\sim$20\% of the total electrons and such non-thermal electrons carry $\sim$50\% of the total electron energy. Note that we do not need to assume cutoff energy nor source volume to estimate the values. Also, these ratios are much larger than what is generally assumed in an electron beam model where beam density is much less than the ambient density (`diluted beam'). Presence of a kappa distribution in a coronal source region may indicate that a significant number of electrons are locally accelerated.


\section{Analysis}
To test the kappa distribution model, we performed imaging spectroscopy for a partially disk-occulted solar flare of 2007 December 31 observed by RHESSI.  Following the work by \cite{krucker10}, our focus goes to the time of HXR peak flux 00:47:42-00:47:50 UT. 

\begin{figure*}
\begin{center}
\includegraphics[width=180mm]{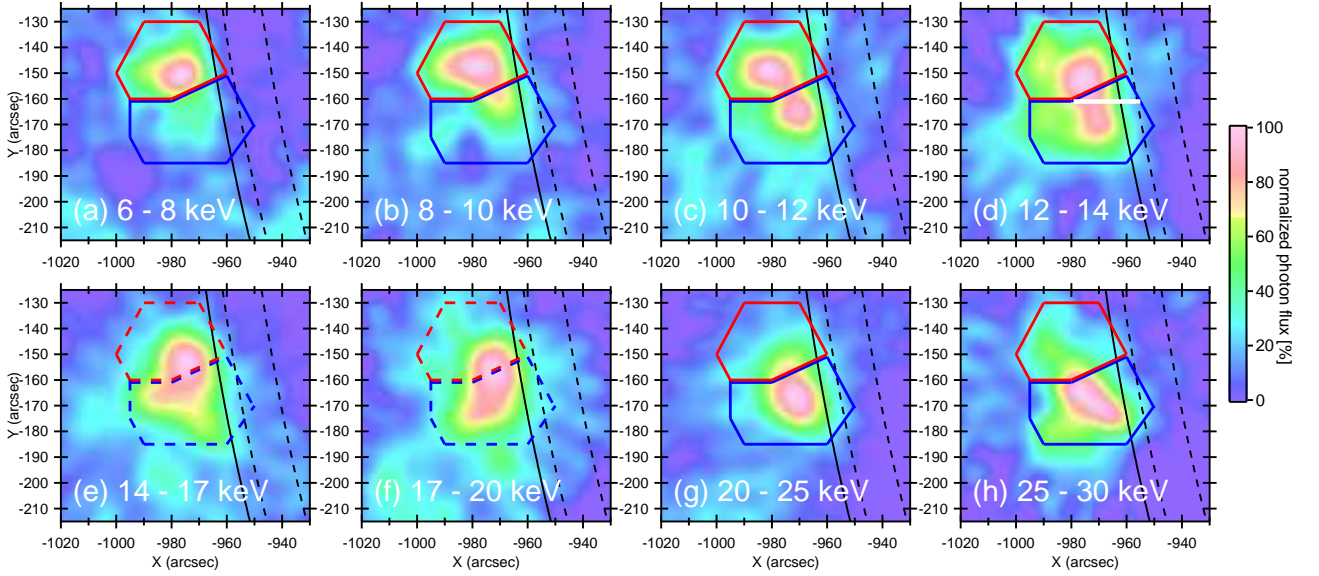}
\caption{
Hard X-ray sources during 00:47:42-00:47:50 UT of 2007 December 31. The images are constructed for eight different energy ranges by the CLEAN algorithm. Subcollimators 3 - 8 are used with natural weighting.  The color-code shows the photon flux normalized by the peak flux within each image. The peak fluxes were (a) 22, (b) 2.1, (c) 0.52, (d) 0.17, (e) 0.12, (f) 0.05, (g) 0.046 and (h) 0.016 [photons cm$^{-2}$ s$^{-1}$ arcsec$^{-2}$].   The blue and red polygons indicate the regions in which the total fluxes were calculated for the spectra in Figure \ref{fig:spectrum}. The dashed polygons in panels (e) and (f) emphasize that the data from these two energy ranges are not used in the analysis. We confirmed that the conclusion of this paper is not sensitive to the choice of subcollimators, construction algorithm and the size/shape of the polygons. The black solid curve gives the location of the photospheric limb.
\label{fig:image}}
\end{center}
\end{figure*}

Figure \ref{fig:image} shows the spatial structure of the HXR sources during the peak time in 8 different energy ranges.  There were mainly two separate sources. The northern source at (X,Y)$\sim$(-980,-150) arcsec was dominated by a low energy ($<$10 keV) X-ray emission (Figure \ref{fig:image}(a,b)) whereas the southern source at (X,Y)$\sim$(-970,-165) arcsec was dominated by a high energy ($>$20 keV) X-ray emission (Figure \ref{fig:image}(g,h)).  The northern source in the 6-8 and 8-10 keV ranges was so bright that the flux of the southern source should be less than what can be calculated in, for example, the blue polygons of Figure \ref{fig:image}(a,b). Conversely, the fluxes in the red polygons of Figure \ref{fig:image}(g,h) would be the upper limits of the northern sources in the 20-25 and 25-30 keV ranges. 

Note that, soon after the HXR peak time, the main thermal loop of the southern source was identified along the limb on the western side of the southern source. Thus, the northern source has been considered as a separate thermal source, although the precise relation between the northern and southern sources remains unclear \citep{krucker10}. Below, we focus on the spectral features of the sources rather than a possible relationship between the two sources.

Figure \ref{fig:image}(c-f) show the details of the HXR sources in the intermediate energy ranges. In the 10-12 and 12-14 keV ranges (Figure \ref{fig:image}c and \ref{fig:image}d, respectively), both northern and southern sources appear together, indicating that both sources had comparable fluxes. In the 14-17 and 17-20 keV ranges (Figure \ref{fig:image}e and \ref{fig:image}f, respectively), only one source can be identified somewhere between the northern and southern sources and it is not clear which of the two sources it belongs to. Because of such unclear nature of the source structure, we took the sum of the fluxes in the dashed polygons and consider it as the upper limit of both northern and southern sources. The sum, however, is essentially the same as the values in the spatially integrated spectrum shown below and so we do not use the data from Figures \ref{fig:image}(e) and \ref{fig:image}(f) in the following analysis. Then, assuming that the red and blue polygons represent the northern and southern sources, respectively, we calculated the total flux within each polygon to be compared with spectrum models. 

It is to be noted that, while we chose the polygons so that the double sources in Figure \ref{fig:image}(c) can be separated, another choice of boundary indicated by the white line in Figure \ref{fig:image}(d) could also be used to better separate the double sources in Figure \ref{fig:image}(d). We found that such modification to the polygons leads to less than 30\% flux change in all energy ranges. Thus, we will use this number as the uncertainty of the measured fluxes.

\begin{figure*}
\begin{center}
\includegraphics[width=180mm]{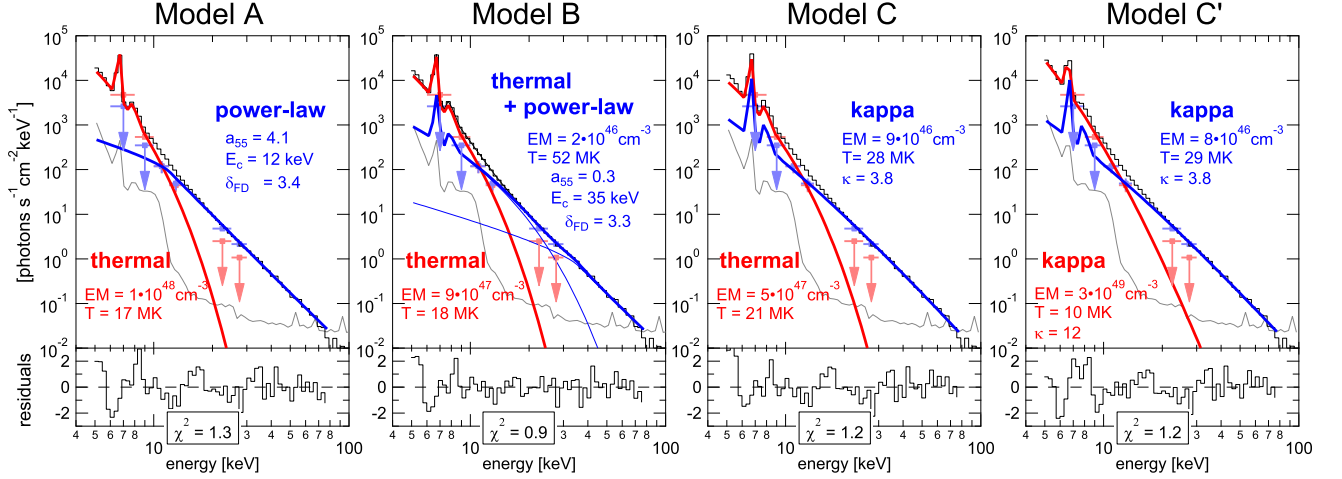}
\caption{
Comparison of the observed spatially integrated photon spectrum (histograms) with modeled distributions (solid curves) as well as the imaging spectroscopy result (light-red and light-blue squares for the northern and southern sources, respectively).  Four different sets of models were used to fit the observed spectrum: (A) thermal + power-law, (B) thermal + thermal + power-law, (C)  thermal + kappa and (C') kappa + kappa. See Tables \ref{tbl-1} and \ref{tbl-2} for the resultant parameter values of the steeper (red) and flatter (blue) component, respectively. The gray curve is the background. The histograms in the lower panels are the residuals of the fit in units of the standard deviation derived from photon statistics. The bremsstrahlung emission formula of power-law, i.e. {\ttfamily f\_thin.pro} is used for Models A and B. This program is implemented as double power-law but we fixed the break energy at 1000 keV to make it a single power-law.  All kappa distributions take into account the line emissions. We used {\ttfamily f\_vth.pro} with the lines-only option and combined it with {\ttfamily f\_thin\_kappa.pro} through a wrapper routine.
\label{fig:spectrum}}
\end{center}
\end{figure*}

Figure \ref{fig:spectrum} shows the result of the imaging spectroscopy. The light-red and light-blue squares indicate the fluxes from red and blue polygons in Figure \ref{fig:image}, and  the fluxes are compared to the spatially integrated photon spectrum (histograms) as well as four different model distributions (colored curves). The steeper and flatter components are evident in the integrated spectrum and we will use thermal, power-law and kappa distributions to represent these two components. To represent the non-thermal tail, Models A and B use the power-law distributions whereas Models C and C' use the kappa distribution.


For the power-law distribution fits used in Models A and B, we used a formula that calculates thin-target Bremsstrahlung X-ray spectrum from a power-law electron flux density distribution (cm$^{-2}$s$^{-1}$keV$^{-1}$). This formula is implemented as {\ttfamily f\_thin.pro} in the SSW/OSPEX software and contains three free parameters $a_{55}$, $\delta_{\rm FD}$ and E$_{\rm c}$.  $a_{55}$ (10$^{55}$ electrons cm$^{-2}$s$^{-1}$) is the normalization factor and is a product of the number density of plasma ions, the flux density of non-thermal electrons, and the volume of the radiating source region. $\delta_{\rm FD}$ is the power-law index of the flux density distribution.  Throughout this paper, however, we use the power-law index $\delta$ of the number density distribution when comparing different models. The two indices can be converted to one another by $\delta$ = $\delta_{\rm FD}$ + 0.5. $E_{\rm c}$ (keV) is the low energy cutoff of the power-law distribution.

For the kappa distribution fits used in Models C and C', we used the SSW/OSPEX procedure {\ttfamily f\_thin\_kappa.pro} but it does not contain line emissions. We added line emissions in the analysis and imposed  that the same values of emission measure (EM) and temperature (T) are used in the kappa distribution and line emissions. It is to be noted that the assumed values for the fits were derived for a Maxwellian distribution, and the temperature inferred under the assumption that electron distribution is Maxwellian may be an overestimate of the actual temperature of a distribution with non-thermal tail \citep{owocki83}. As for the spectral index, the program {\ttfamily f\_thin\_kappa.pro} assumes a kappa distribution for the electron density and the spectral index $\kappa$ can be converted to $\delta$ by $\delta$ = $\kappa$ + 0.5.


Figure \ref{fig:spectrum} Model A uses the thermal (red curve) and power-law (blue curve) distributions for the steeper and flatter components, respectively. The light-red and light-blue squares are  consistent with the red and blue curves, respectively, indicating that the northern source was producing the steeper component whereas the southern source was producing the flatter component. Note that a range of values for the low-energy cutoff $E_{\rm c}$ fit the data equally well. We found that the highest $E_{\rm c}$ that still fits the data is 16 keV with $\chi^2$ = 1.0 (not shown). However,  this model underestimates the flux in the 10-12 keV range by $46^{+20}_{-11}$ \% of what was measured in the image (light-blue square). Then, we imposed $E_{\rm c}$ to be in the range 11 $< E_{\rm c} <$ 13 keV in order to look for a solution consistent with the imaging spectroscopy result. We found that, as shown in Figure \ref{fig:spectrum} Model A,  $E_{\rm c}$ = 12 keV would best fit the data,  although  $\chi^2$ became relatively large ($\chi^2$ = 1.3).

Figure \ref{fig:spectrum} Model B uses the thermal (red curve) and combination of thermal and power-law (blue curve) distributions for the steeper and flatter components, respectively. This is basically the same as Model A, but an additional thermal distribution is introduced to account for the possible core component of the southern source (blue curve).  A similar model is used by \cite{caspi10}. This set of distributions can also fit the data nicely ($\chi^2$ = 0.9) including the 10-12 keV range, suggesting that a core distribution could have existed in the southern source. Again, a range of values for the low-energy cutoff $E_{\rm c}$ fit the data equally well and we chose the highest $E_{\rm c}$ that still fits the data ($E_{\rm c}$ = 35 keV). Note that the model curves (generated to fit the integrated photon spectrum) have a higher energy resolution so that the modeled line emissions partially exceed the fluxes obtained from images in the 6-8 and 8-10 keV ranges. We confirmed that the modeled values averaged over the same energy ranges are consistent with the values from the imaging spectroscopy. In Model B, the thermal core component has a relatively large temperature of $\sim$52 MK and this is comparable to the temperature $\sim$44 MK of a super-hot coronal source reported by \cite{caspi10}.

Figure \ref{fig:spectrum} Model C uses the thermal (red curve) and kappa distributions (blue curve) for the steeper and flatter components, respectively. Although the reduced $\chi^2$ is slightly larger ($\chi^2$ = 1.2), this set of models can also represent the data nicely including the 10-12 keV range. Again, the modeled line emissions partially exceed the fluxes obtained from images in the 6-8 and 8-10 keV ranges, but we confirmed that the modeled values averaged over the same energy ranges are consistent with the imaging spectroscopy. The number of free parameters, 5, is still the same as that of Model A (thermal + power-law) whereas Model B (thermal + thermal + power-law) needed 7 parameters to have a thermal core distribution in the flatter component (blue curve). 

Figure \ref{fig:spectrum} Model C' uses the kappa distributions for both steeper and flatter components. It is evident that the non-thermal tail of the steeper component is still below the upper limits (light-red arrows). The slope is quite soft ($\kappa\sim$12), however, indicating that the steeper component (northern source) was mostly thermal. Note also that the core temperature of the steeper component 10 MK is reduced by 50\% compared to the temperature from the thermal fit, 21 MK. While such low temperature of the steeper component is still comparable to the temperature 15 MK measured by GOES during the same interval (00:47:42-00:47:50), the emission measure obtained by the fit was unrealistically high, 2.5$\times$10$^{49}$cm$^{-3}$, and it is an order of magnitude higher than what was measured by GOES, 1.5$\times$10$^{48}$cm$^{-3}$. Therefore, this set of two kappa distributions is not favorable for representing the observation.

\begin{figure}
\begin{center}
\includegraphics[width=90mm]{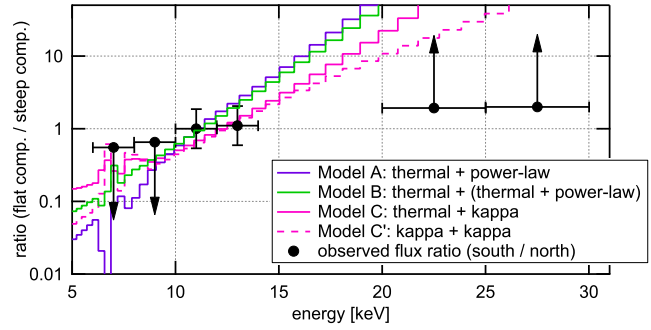}
\caption{
The ratio of flatter component (blue curves in Figure \ref{fig:spectrum}) to steeper component (red curves in Figure \ref{fig:spectrum}) derived by four different sets of models (histograms). The ratio is compared to the flux ratio (filled circles) derived from images (Figure \ref{fig:image}). We did not use the images of 14-17 and 17-20 keV ranges because of the unclear nature of the sources. 
\label{fig:ratio}}
\end{center}
\end{figure}

To better visualize the above comparisons, Figure \ref{fig:ratio} uses the model curves in the 5-31 keV range and takes the ratio of flatter component (blue curves in Figure \ref{fig:spectrum}) to steeper component (red curves in Figure \ref{fig:spectrum}). The ratios are compared to the flux ratios from images (black squares with error bars). The modeled values are averaged over the energy ranges of the images. It is evident that all models are consistent with the imaging spectroscopy. The upper limits in the $<$ 10 keV energy range indicate that, in this energy range, X-ray emission from thermal plasma in the southern source could not have been detected by RHESSI if its flux is less than 65\% of the flux from the northern source. 

\begin{figure}
\begin{center}
\includegraphics[width=90mm]{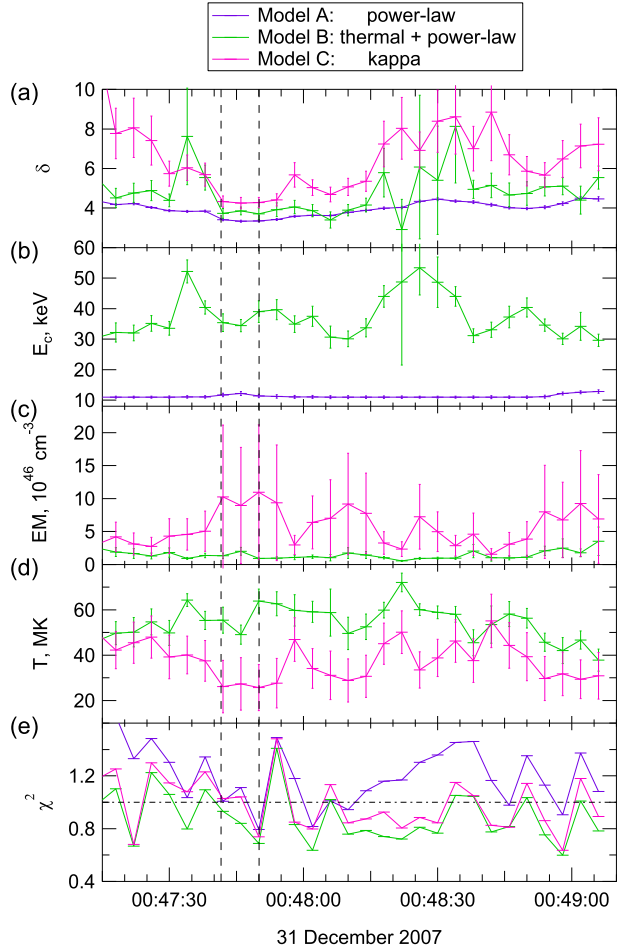}
\caption{
Time variations of the fitted parameters for the flatter component (i.e. southern source) obtained by Model A (purple), Model B (green) and Model C (pink).  From top to bottom are (a) the spectral index $\delta$ and (b) the low-energy cutoff $E_c$ of the non-thermal tail of the southern source, (c) the emission measure EM and (d) temperature of the thermal core of the southern source, and (e) the reduced $\chi^2$ of the model fits. The result by Model C' is not shown because the error ranges (sigma levels) of EM and T exceeded the parameter values. While the peak spectra in Figure \ref{fig:spectrum} were taken during the 8 s interval of 00:47:42-00:47:50 UT (indicated by the dashed lines), we fitted the data every 4s over the nearly two-minutes interval of 00:47:15 - 00:49:10 UT in this figure.
\label{fig:TVnew}}
\end{center}
\end{figure}

Figure \ref{fig:TVnew} examines the time variations of the fitted parameters for the flatter component (i.e. southern source). While the peak time spectra in Figures \ref{fig:spectrum} were taken during the 8 s interval of 00:47:42-00:47:50 UT, we fitted the data every 4 s in Figure \ref{fig:TVnew}. The spectral indices $\delta$ from all models show soft-hard-soft variation as already reported by \citep{krucker10}. The low-energy cutoff $E_c$ of Model B (power-law with thermal core) is  systematically larger than that of Model A (power-law without thermal core) because of the presence of the thermal core distribution in the flatter component (southern source). As for the parameters of the core distribution, the emission measure $EM$ and the temperature $T$ of Model B (thermal+ power-law) are systematically lower and higher, respectively, than those of Model C (kappa). Therefore, Model B suggests presence of a superhot thermal core distribution in the flatter component (southern source) whereas Model C suggests presence of larger number of non-thermal electrons. The reduced $\chi^2$ fluctuated around $\sim$ 1, indicating that Models A, B and C fitted the data fairly well. The averages in the shown interval (00:47:15 - 00:49:10 UT) are Model A: 1.22, Model B: 0.88 and Model C: 0.94.

\section{Discussion}

Let us now discuss implications of the results based on Models A, B and C for the peak flux interval (00:47:42 - 00:47:50 UT). We will particularly discuss non-thermal fractions of electron number/energy densities in the southern source. The estimated non-thermal fractions are summarized in Table 3.

Model A (thermal + power-law) assumes that the southern source (`above-the-loop' coronal source) contains  a negligible amount of thermal electrons and uses the power-law with no thermal core to represent the flatter component. To estimate the number density of non-thermal electrons (`instantaneous' density), we assumed a source volume $V = 8\times10^{26}$ cm$^{ 3}$ \citep{krucker10} and applied the formula by \cite{lin74} to the power-law part of the photon spectrum. To estimate the number density of thermal electrons (`ambient density'), we assumed that the ambient environment should be similar to that of the nearby thermal source. Following the derivation by \cite{krucker10}, the ratio $N_{\rm nt}/N_{\rm th}$ can be expressed as
\begin{equation}
\frac{N_{\rm nt}}{N_{\rm th}} = 0.05 \left( \frac{N^{\rm upper}_{\rm th}}{N_{\rm th}} \right)^2 \left(\frac{E_{\rm c}}{12 {\rm keV}} \right)^{-2.9}
\end{equation}
where $N_{\rm th}^{\rm upper}$ = 8$\times$10$^9$ cm$^{-3}$ is the upper limit of the ambient density and we used $\gamma$ = 4.4, flux at 50 keV of 0.16 ph s$^{-1}$ cm$^{-2}$ keV$^{-1}$ and the low-energy cutoff $E_{\rm c}$ = 12 keV. If we use $N_{\rm th} = N_{\rm th}^{\rm upper}$, we obtain $R_N = N_{\rm nt}/N_{\rm tot} \sim$ 0.05. If we use the best estimate of $N_{\rm th}$ = 2$\times$10$^9$ cm$^{-3}$ \citep{krucker10}, we obtain $R_N \sim$ 0.44. Similarly, the ratio $\varepsilon_{\rm nt}/\varepsilon_{\rm th}$ can be expressed as
\begin{equation}
\frac{\varepsilon_{\rm nt}}{\varepsilon_{\rm th}} = 0.47 \left( \frac{N^{\rm upper}_{\rm th}}{N_{\rm th}} \right)^2 \left(\frac{E_{\rm c}}{12 {\rm keV}} \right)^{-2.9}
\end{equation}
where we assumed that the temperature of the ambient plasma is 22 MK \citep{krucker10}. Then, if we use $N_{\rm th} = N_{\rm th}^{\rm upper}$, we obtain $R_\varepsilon \sim$ 0.32. If we use the best estimate of $N_{\rm th}$ = 2$\times$10$^9$ cm$^{-3}$, we obtain $R_\varepsilon \sim$ 0.88. Because of this large fraction of non-thermal electrons, Model A implies that the non-thermal electrons are not simply a tail on the thermal distribution and that electrons are accelerated locally in the southern source (i.e. `above-the-loop' coronal source). However, Model A resulted in a relatively large $\chi^2$ ($\sim$1.3 at the flux peak time and $\sim$1.2 on average). Therefore, we explored other possible models as described below.

Model B (thermal + thermal + power-law) assumes that the southern source (`above-the-loop' coronal source) contains a significant amount of thermal electrons and uses the power-law with a hot (52 MK) thermal core to represent the flatter component. To estimate the non-thermal fraction of electron number density, we can use the obtained normalization factor a$_{\rm 55}$ ($\sim$ 0.3 $\times$ 10$^{55}$ cm$^{-1}$s$^{-1}$) because it is actually a product of the number density of plasma ions, the flux density of non-thermal electrons, and the volume of the radiating source region. Using the plasma density 4.5 $\times$ 10$^9$ cm$^{-3}$ (Table \ref{tbl-2}) and an assumed source volume of 8$\times$10$^{26}$cm$^3$ \citep{krucker10}, the electron flux density is estimated to be  8$\times$ 10$^{17}$ cm$^{-2}$s$^{-1}$. Based on the low-energy cutoff energy E$_{\rm c}$ = 35 keV,  the mean speed of the accelerated electrons can be estimated on the order of 10$^{10}$ cm/s and the number density of the non-thermal electrons in the southern X-ray source is estimated to be 8$\times$ 10$^{7}$cm$^{-3}$. This is only 2\% of the thermal electron density. 


Such small fraction of non-thermal electron density is consistent with an electron beam scenario in which electrons are accelerated above the hot flare loops and stream through the source region to produce a super-hot thermal plasma.  Model B is also consistent with our imaging spectroscopy result especially in the 10-12 keV range, justifying our assumption that a thermal core distribution may have existed in the southern X-ray source (`above-the-loop' coronal source). In fact, Model B gives the least $\chi^2$ (= 0.9) and, as such, seems to be a plausible model. 

A caveat is that, because it contains an additional thermal distribution, the number of free parameters, 7, is relatively larger compared to 5 in Models A and C. In general, a larger number of free parameters contributes to decreasing the reduced $\chi^2$. Therefore, Model B may have resulted in the lower $\chi^2$ partly because of the smaller number of parameters, although the number of free parameters alone does not explain the $\chi^2$ difference. It is also to be noted that a range of values for $E_{\rm c}$ fit the data equally well and the highest value of $E_{\rm c}$ that still fits the data has been used in Model B. As such, the above estimation of the fraction of non-thermal electrons, 2\%, should only be considered as a lower-limit.

In both power-law Models A and B, we needed to assume the source volume  $V$ as well as the low-energy cutoff $E_{\rm c}$ to estimate the fraction of non-thermal electrons. Note that we can only obtain a very rough estimate of $V$ and the estimated fraction of non-thermal electrons can be sensitive to the choice of $E_{\rm c}$.  Then, we consider the kappa distribution as an alternative  because it contains a thermal core component that seamlessly extend to a power-law distribution.  The kappa distribution allows us to estimate the fraction of non-thermal electron density/energy without assuming the source volume (Section \ref{sec:kappa}).

In Model C (thermal + kappa),  the ratio of non-thermal electron density to the total electron density in the southern source was $R_{\rm N}$ = 0.20$^{+0.01}_{-0.01}$, and the ratio of non-thermal energy to the total electron energy in the southern source was $R_{\rm \varepsilon}$ = 0.52$^{+0.03}_{-0.02}$. It is to be emphasized that the non-thermal fractions of number/energy densities have been derived less ambiguously than Model A and that the result does not invoke the possibility of non-thermal electrons outnumbering thermal electrons. On the other hand, the derived estimate of $R_{\rm N}$ = 0.20$^{+0.01}_{-0.01}$ is much larger than $R_{\rm N} \sim$ 0.02 of Model B as derived above. This implies that not all electrons are thermalized in the southern (`above-the-loop') source region and that there may have been local acceleration of electrons in this region. 

If we assume the source volume of 8$\times$10$^{26}$cm$^3$ \citep{krucker10}, the total density of the southern source $N_{\rm tot}$ can be estimated as (1.0 $\pm$ 0.9) $\times$10$^{10}$ cm$^{-3}$. Within the error range, the estimated density is consistent with the upper limit of $8\times10^9$ cm$^{-3}$ derived by \cite{krucker10}. The estimated density is also consistent with what was estimated in Model B, (4.5 $\pm$ 2.2) $\times$10$^{9}$ cm$^{-3}$ (Table \ref{tbl-2}).

As for temperature, the `above-the-loop' region (i.e. southern source) had a temperature of 28 $\pm$ 9 MK. This is $\sim$ 1.3 times larger than the temperature 21 $\pm$ 0.4 MK of the nearby thermal source but is $\sim$ 0.6 times the temperature of the super-hot component discussed in Model B  or \cite{caspi10}. We speculate that the released magnetic field energy was converted to both thermal and non-thermal energies of electrons but a significant fraction ($R_{\rm \varepsilon}\sim$ 0.5) went to non-thermal electrons so that the temperature did not increase considerably.

As for the effective plasma beta $\beta$, \cite{krucker10} estimated the magnetic field strength $|B|$ to be 30 - 50 G and derived $\beta$ between $\sim$0.005 and $\sim$0.02 for a pre-flare plasma  with the density 2 $\times$ 10$^9$ cm $^{-3}$ and the temperature 2 MK. They argued that the preflare thermal plasma could be replaced with non-thermal electrons (the low-energy cutoff at 16 keV) so that the effective plasma beta becomes $\beta \sim$ 1. If we use the density $\sim$10$^{10}$ cm$^{-3}$ and the kappa temperature $\sim$ 28 MK as derived from Model C, the effective plasma beta falls between $\sim$ 1 and $\sim$ 3. 

From the spectral fit, we obtained $\kappa\sim$3.8 which leads to $\delta\sim$4.3 for the density distribution $F(E)\propto E^{-\delta}$ (see Equation (2)). This is somewhat softer compared to $\delta\sim$3.9 derived from Model A (thermal +  power-law) and $\delta\sim$3.8 derived from Model B (thermal + thermal + power-law). The $\kappa$-value is not too large, however, and the kappa distribution is far from a single Maxwellian. Thus, an electron acceleration theory still needs to reproduce a power-law tail for this event. Note again that the power-law index $\delta$ can be converted from $\delta_{\rm FD}$ of the flux density distribution used in {\ttfamily f\_thin.pro} (Models A \& B; $\delta = \delta_{\rm FD}$ + 0.5) and from $\kappa$ of the number density distribution used in {\ttfamily f\_thin\_kappa.pro} (Models C \& D; $\delta = \kappa$ + 0.5).

It is to be mentioned that Model C also has caveats and disadvantages. First, $\chi^2$ was relatively larger ($\chi^2\sim$ 1.2 at the peak flux interval 00:47:42 - 00:47:50.  The relatively large $\chi^2$ may imply that the kappa distribution is still not the best functional form to represent the HXR spectrum from the southern source. However, the reduced $\chi^2$ averaged over the larger interval 00:47:15 - 00:49:10 was smaller $\sim$ 0.9 and this is comparable to the average $\chi^2 \sim$ 0.9 of Model B. The time variation of $\chi^2$ was also similar between Models B and C (Figure \ref{fig:TVnew}). Thus,  Model C is as acceptable as Model B in terms of spectral fitting. Second, the spectral index at the time of peak flux $\delta\sim$4.3 is even more inconsistent with that deduced from the radio observations ($\delta\sim$3.4; \cite{krucker10}) than that deduced from the other models ($\delta\sim$3.9 in Model A and $\delta\sim$3.8 in Model B). However, the radio emission represents electrons with energies larger than $\sim$ 100 keV whereas our analysis was made in the $<$ 100 keV range. To further understand if these disadvantages are common in  other solar flare events, it is important to test the kappa distribution in a larger number of solar flare events.

\section{Conclusion}

The kappa distribution does not require a low-energy cutoff $E_{\rm c}$ to represent non-thermal electrons, and the thermal core component can seamlessly extend to a power-law distribution. Furthermore, the non-thermal fractions of electron number/energy densities can be uniquely estimated because they are functions of the power-law index $\kappa$ only. While \cite{kasparova09} applied the kappa distribution to loop-top coronal sources, we examined the kappa distribution model in an unusually bright `above-the-loop' coronal source obtained on 2007 December 31 \citep{krucker10}. For comparison, we also examined the conventional power-law models with and without a thermal core distribution in the source.

Model A, the power-law with no thermal core component, was consistent with the imaging spectroscopy result when we chose $E_{\rm c}$ = 12 keV, although the reduced $\chi^2$ was relatively  large ($\sim$1.2 on average). This model implies that non-thermal electrons can outnumber thermal electrons.

Model B, the power-law  combined with a super-hot (52 MK) thermal core component, could fit the observed spectrum well ($\chi^2 \sim$ 0.9 on average) and was consistent with the imaging spectroscopy. This model implies that at least 2\% of the source electrons carried non-thermal energies. 

However, both Models A and B  require a low-energy cutoff $E_{\rm c}$ to represent the non-thermal tail, and the estimates of the electron number/energy densities can be sensitive to the choice of $E_{\rm c}$. Furthermore,  a source volume $V$ had to be assumed for the estimates but we can only obtain a rough estimate of $V$.

Thus, we examined Model C (the kappa distribution model). We found that it can fit the observed spectrum well ($\chi^2 \sim$ 0.9 on average) and is consistent with the imaging spectroscopy result. Without assuming the source volume $V$ and the lower-energy cutoff $E_{\rm c}$, we estimated that a moderate fraction (20\%) of the source electrons had non-thermal energies and carried 52\% of the total electron energy in the `above-the-loop' coronal source region. The temperature was 28 MK and the power-law index of the electron density distribution was  -4.3. It would be important to examine a larger number of events in order to verify the generality of the kappa distribution model.


\acknowledgments

We acknowledge helpful comments by the anonymous referee and L. Glesener, T. D. Phan and M. Hoshino. MO was supported by NASA grant NNX08AO83G at UC Berkeley. PSH was supported by NASA grant NAS5-98033. RPL was supported by the WCU Grant (R31-10016) funded by the Korean Ministry of Education, Science and Technology.

\clearpage

\begin{center}
\begin{deluxetable}{lcccc}
\tabletypesize{\scriptsize}
\tablecaption{Fit result of the steeper component (colored red in Figure \ref{fig:spectrum}\label{tbl-1}) }
\tablewidth{0pt}
\tablehead{
\colhead{fit parameters } &
\colhead{Model A} &
\colhead{Model B} &
\colhead{Model C} &
\colhead{Model C'} \\
\colhead{            }& 
\colhead{thermal } & 
\colhead{thermal } & 
\colhead{thermal } & 
\colhead{kappa   } 
}

\startdata
emission measure EM, cm$^{-3}$&
$(1.2 \pm 0.1) \times 10^{48}$ & 
$(8.6 \pm 1.0) \times 10^{47}$ &
$(4.8 \pm 0.2) \times 10^{47}$ &
$(2.5 \pm 1.0) \times 10^{49}$ \\

total density\tablenotemark{a} N$_{\rm tot}$, cm$^{-3} $&
$(3.9 \pm 1.3) \times 10^{10}$ & 
$(3.3 \pm 1.1) \times 10^{10}$ &
$(2.4 \pm 0.6) \times 10^{10}$ &
$(1.8 \pm 1.1) \times 10^{11}$ \\

temperature\tablenotemark{b} T, MK             &
$ 17 \pm 0.5 $ &
$ 18 \pm 0.6 $ &
$ 21 \pm 0.4 $ &
$ 10 \pm 0.7 $ \\

power-law index\tablenotemark{c} $\delta$ &
$ -  $ &
$ -  $ &
$ -  $ &
$ 12 \pm 0.8 $ 

\enddata

\tablenotetext{a}{
The density is derived from the emission measure by assuming a source volume of $\sim 8 \times 10^{26} $cm$^{3}$ \citep{krucker10}.
}
\tablenotetext{b}{
The kappa temperature $k_{\rm B}T_\kappa$ is used for the kappa distribution fit.
}
\tablenotetext{c}{
The power-law index $\delta (= \kappa + 0.5) $ is for the density distribution $F(E)\propto E^{-\delta}$.
}
\end{deluxetable}
\end{center}

\begin{center}
\begin{deluxetable}{lccccc}
\tabletypesize{\scriptsize}
\tablecaption{Fit result of the flatter component (colored blue in Figure \ref{fig:spectrum}\label{tbl-2})}
\tablewidth{0pt}
\tablehead{
\colhead{fit parameters } &
\colhead{Model A} &
\multicolumn{2}{c}{Model B} &
\colhead{Model C} &
\colhead{Model C'} \\
\colhead{            }& 
\colhead{power-law } & 
\colhead{thermal} &
\colhead{power-law} &
\colhead{kappa   } & 
\colhead{kappa   } 
}

\startdata
emission measure EM, cm$^{-3} $&
$ - $ &
$(1.6 \pm 0.3) \times 10^{46}$ &
$ -                          $ &
$(8.6 \pm 6.2) \times 10^{46}$ & 
$\sim 8        \times 10^{46}$ \\

total density N$_{\rm tot}$, cm$^{-3} $&
$ - $ &
$(4.5 \pm 2.2) \times 10^{ 9}$ &
$ -                          $ &
$(1.0 \pm 0.9) \times 10^{10}$ &
$\sim 1        \times 10^{ 9}$ \\

temperature T, MK&
$ - $ &
$ 52 \pm  4$ &
$ -        $ &
$ 28 \pm  9$ &
$ 29 \pm 18$ \\

power-law index $\delta        $&
$ 3.9 \pm 0.04 $ &
$ -           $ &
$ 3.8 \pm 0.2 $ &
$ 4.3 \pm 0.2 $ &
$ 4.2 \pm 0.4 $
\enddata

\tablecomments{
Same format as Table \ref{tbl-1}. An error range (sigma level) is not shown when it exceeded the parameter value. The power-law index $\delta$ is for the density distribution $F(E)\propto E^{-\delta}$ and  can be converted from $\delta_{\rm FD}$ of the flux density distribution used in {\ttfamily f\_thin.pro} (Models A \& B; $\delta = \delta_{\rm FD}$ + 0.5) and from $\kappa$ of the number density distribution used in {\ttfamily f\_thin\_kappa.pro} (Models C \& D; $\delta = \kappa$ + 0.5).
}
\end{deluxetable}
\end{center}

\clearpage

\clearpage

\begin{center}
\begin{deluxetable}{lccc}
\tabletypesize{\scriptsize}
\tablecaption{Non-thermal fractions of electron number/energy densities in the southern source}
\tablewidth{0pt}
\tablehead{
\colhead{type of} &
\colhead{Model A} &
\colhead{Model B} &
\colhead{Model C} \\
\colhead{density} &
\colhead{(power-law without thermal core)} &
\colhead{(power-law with thermal core)} &
\colhead{(kappa)} 
}

\startdata
number density&
 44 \% ($>$5 \%) &
 $>$2 \% &
 20 $\pm$ 1 \% \\

energy density&
 88 \% ($>$32 \%) &
 $>$12 \% &
 52$^{+3}_{-2}$ \% 
\enddata


\tablecomments{
See text for details of the estimations. For Model A, the lower limits of the number/energy densities (shown in parenthesis) are based on the upper limit of the estimated number density of the ambient electrons \citep{krucker10}. For Model B, the inequalities ($>$) are based on the fact that a range of values for the low-energy cutoff $E_c$ fit the data equally well and we used the cases of the highest $E_c$. The result by Model C' is not shown because the error ranges (sigma levels) of emission measure EM and temperature T exceeded the parameter values.  }
\end{deluxetable}
\end{center}

\end{document}